\documentclass[graybox]{svmult}

\usepackage{mathptmx}       
\usepackage{helvet}         
\usepackage{courier}        
\usepackage{type1cm}        

\usepackage{makeidx}         
\usepackage{graphicx}        
\usepackage{multicol}        
\usepackage[bottom]{footmisc}

\begin{document}

\title*{Group classification of variable coefficient KdV-like equations}
\author{Olena Vaneeva}
\institute{Olena Vaneeva \at Institute of Mathematics of the
National Academy of Sciences of Ukraine, 3 Tereshchenkivs'ka Street,
01601 Kyiv-4, Ukraine. \email{vaneeva@imath.kiev.ua}}
\maketitle

\abstract{The exhaustive group classification of the class of KdV-like equations with
time-dependent coefficients $u_t+uu_x+g(t)u_{xxx}+h(t)u=0$ is carried out using equivalence based approach.
A simple way for the construction of exact solutions  of KdV-like equations
using equivalence transformations is described.}

\section{Introduction}
A number of physical processes are modeled by generalizations of the well-known
equations of mathematical physics such as, e.g., the KdV and mKdV equations,
the Kadomtsev--Petviashvili  equation, which contain time-dependent coefficients.
That is why last decade these equations do attract attention of researchers.
A number of the papers devoted to the study of variable coefficient KdV or mKdV equations with time-dependent coefficients
were commented in~\cite{Popovych&Vaneeva2010}.
In the majority of papers  the results were
obtained mainly for the equations which are reducible to the standard KdV or mKdV equations by point transformations.
Unfortunately equivalence properties are neglected usually and
finding of exact solutions is reduced to
complicated calculations of systems involving a number of unknown functions
using computer algebra packages. It is shown in~\cite{Popovych&Vaneeva2010,Vaneeva2012} that
 the usage of equivalence transformations allows one to obtain the results in a much simpler way.

In this paper this fact is reaffirmed via presentation the correct group classification of a class of variable coefficient
KdV equations using equivalence based approach. Namely,
we investigate Lie symmetry properties and exact solutions of variable coefficient
KdV equations of the  form
\begin{equation}\label{vc_mKdV}
u_t+uu_x+g(t)u_{xxx}+h(t)u=0,
\end{equation}
where $g$ and $h$ are arbitrary smooth functions of the variable $t$, $g\neq0.$
It is shown in Section~2 that using equivalence transformations the function $h$ can be always set to the zero value and therefore
the form of $h$ does not affect results of group classification.
The group classification of class~(\ref{vc_mKdV}) with $h=0$ is carried out in~\cite{Popovych&Vaneeva2010}.
So, using the known classification list and equivalence transformations we present group classification of the initial
class~(\ref{vc_mKdV}) without direct calculations.

An interesting property of class~(\ref{vc_mKdV})  is that it is normalized,
i.e., all admissible point transformations within this class are generated by transformations
from the corresponding equivalence groups.
Therefore, there are no additional equivalence transformations between cases of the classification list,
which is constructed using the equivalence relations associated with the corresponding equivalence group.
In other words, the same list represents the group classification result for the corresponding class
up to the general equivalence with respect to point transformations.

Recently the authors of~\cite{john10b} obtained a partial group classification of class~(\ref{vc_mKdV})
(the notation $a$ and $b$ was used there instead of $h$ and $g$, respectively).
The reason of failure was neglecting an opportunity to use equivalence transformations.
This is why only some cases of Lie symmetry extensions were found, namely the cases with $h={\rm const}$, $h=1/t$ and $h=2/t$.

In fact the group classification problem
for class~(\ref{vc_mKdV})  up to its
equivalence group is already solved since this class is reducible to class~(\ref{vc_mKdV}) with $h=0$
whose group classification is carried out in~\cite{Popovych&Vaneeva2010}.
Using the known classification list and equivalence transformations we present group classifications of
class~(\ref{vc_mKdV})
without the simplification of both equations admitting extensions of Lie symmetry algebras and these
algebras themselves by equivalence transformations.
The extended classification list can be useful for applications and convenient to be compared
with the results of~\cite{john10b}.

Note that in~\cite{Gungor&Lahno&Zhdanov2004,Magadeev1993} group classifications for more general classes that include
 class~(\ref{vc_mKdV}) were carried out. Nevertheless those results obtained up to very wide equivalence group
  seem to be inconvenient to derive group classification for class~(\ref{vc_mKdV}).

\section{Equivalence transformations}
An important step under solving a group classification problem is the construction of the equivalence group
of the class of differential equations under consideration.
The usage of transformations from the related equivalence group often gives an opportunity to essentially simplify a
group classification problem and to present the final results in a closed and concise form.
Moreover, sometimes this appears to be a crucial point in
the exhaustive solution of such problems~\cite{IPS2007a,Vaneeva2012,VJPS2007,VPS_2009}.

There exist several kinds of equivalence groups.
The \emph{usual equivalence group} of a class of differential equations consists of the nondegenerate point transformations
in the space of independent and dependent variables and arbitrary elements of the class
such that the transformation components for the variables do not depend on arbitrary elements
and each equation from the class is mapped by these transformations to equations from the same class.
If any point transformation between two fixed equations from the class belongs to its (usual) equivalence group then
this class is called \emph{normalized}.
See theoretical background on normalized classes in~\cite{Popovych2006c,Popovych&Kunzinger&Eshraghi2010}.

We find the equivalence
group $G^\sim_{1}$ of class~(\ref{vc_mKdV}) using the results obtained in~\cite{Popovych&Vaneeva2010} for
more general class of variable coefficient KdV-like equations.
Namely, in~\cite{Popovych&Vaneeva2010} a hierarchy of normalized subclasses
of the general third-order evolution equations was constructed.
The equivalence group for normalized class of variable coefficient KdV
equations
\begin{equation}\label{EqvcmKdV}
u_t+f(t)uu_x+g(t)u_{xxx}+h(t)u+(p(t)+q(t)x)u_x+k(t)x+l(t)=0,
\end{equation}
as well as criterion of reducibility of equations from this class to the standard KdV equation were found therein.

The equivalence group $G^\sim$ of class~(\ref{EqvcmKdV}) consists of the transformations
\begin{equation}\label{EqvcKdVEquivGroup}
\tilde t=\alpha(t),\quad
\tilde x=\beta(t)x+\gamma(t),\quad
\tilde u=\theta(t)u+\varphi(t)x+\psi(t),\quad
\end{equation}
where $\alpha$, $\beta$, $\gamma$, $\theta$, $\varphi$ and $\psi$ run through the set of smooth functions of~$t$, $\alpha_t\beta\theta\ne0$.
The arbitrary elements of~(\ref{EqvcmKdV}) are transformed as follows
\begin{eqnarray}\label{EqvcKdVEquivGroupArbitraryElementTrans1}
&\displaystyle\tilde f=\frac{\beta}{\alpha_t\theta}f, \quad
\tilde g=\frac{\beta^3}{\alpha_t}g, \quad
\tilde h=\frac1{\alpha_t}\left(h-\frac\varphi\theta f-\frac{\theta_t}\theta\right),
\\
&\displaystyle\tilde q=\frac1{\alpha_t}\left(q-\frac\varphi\theta f+\frac{\beta_t}\beta\right),\quad \tilde p=\frac1{\alpha_t}\left(\beta p-\gamma q+\frac{\gamma\varphi-\beta\psi}\theta f+\gamma_t-\gamma\frac{\beta_t}\beta\right), \\
&\displaystyle\label{EqvcKdVEquivGroupArbitraryElementTrans3}\tilde k=\frac1{\alpha_t\beta}\left(\theta k-\varphi\alpha_t\tilde h-\varphi_t\right), \quad
\tilde l=\frac1{\alpha_t}\left(\theta l-{\gamma}{\alpha_t}\tilde k-\psi\alpha_t\tilde h-\varphi p-\psi_t\right).
\end{eqnarray}
We also adduce the criterion of reducibility of~(\ref{EqvcmKdV}) to the standard KdV equation.
\begin{proposition}[\cite{Popovych&Vaneeva2010}]
An equation of form~(\ref{EqvcmKdV}) is similar to the standard (constant coefficient) KdV equation
if and only if its coefficients satisfy the condition
\begin{equation}\label{EqvcKdVEquivToKdV}
s_t=2gs^2-3qs+\frac fgk, \quad\mbox{where}\quad s:=\frac{2q-h}g+\frac{f_tg-fg_t}{fg^2}.
\end{equation}
\end{proposition}

Class~(\ref{vc_mKdV}) is a subclass of class~(\ref{EqvcmKdV}) singled out by the conditions $f=1$ and $p=q=k=l=0.$
Substituting these values of the functions $f, p, q, k$ and $l$ to~(\ref{EqvcKdVEquivToKdV}) we obtain
the following assertion.

\begin{corollary}
An equation from class~(\ref{vc_mKdV}) is reduced to the standard KdV equation by a point transformation if and only if there exist a constant $c_0$   and $\varepsilon\in\{0,1\}$ such that
\begin{equation}\label{EqvcKdVEquivToKdV2}
h=\frac{\varepsilon}2\frac{g}{\int\!g\, dt+c_0}-\frac{g_t}g.
\end{equation}
\end{corollary}

As class~(\ref{EqvcmKdV}) is normalized~\cite{Popovych&Vaneeva2010}, its equivalence group $G^\sim$ generates
the entire set of admissible (form-preserving) transformations for this class.
Therefore, to describe the set of admissible transformations for class~(\ref{vc_mKdV})
we should set $\tilde f=f=1,$ $\tilde p=p=\tilde q=q=\tilde k=k=\tilde l=l=0$
in~(\ref{EqvcKdVEquivGroupArbitraryElementTrans1})--(\ref{EqvcKdVEquivGroupArbitraryElementTrans3})
and solve the resulting equations with respect to transformation parameters.
It appears that class~(\ref{vc_mKdV}) admits generalized extended equivalence group and it is normalized in generalized sense only.

Summing up the above consideration, we formulate the following theorem.
\begin{theorem}
The generalized extended equivalence group~$\hat G^{\sim}_1$ of class~(\ref{vc_mKdV}) consists of the transformations
\[
\displaystyle\tilde t=\alpha,\ \
\tilde x=\beta x+\gamma,\ \ \tilde u=\lambda(\beta u+\beta_t x+\gamma_t),\ \
\tilde h=\lambda\, h-2\lambda\frac{\beta_t}{\beta}-\lambda_t, \ \
\tilde g=\beta^3\lambda\, g.
\]
Here $\alpha$ is an arbitrary smooth
function of $t$ with $\alpha_t\neq0,$ $\beta=(\delta_1\int e^{-\int h dt} dt+\delta_2)^{-1}$, $\gamma=
\delta_3\int \beta^2e^{-\int h dt} dt+\delta_4$; $\delta_1,\dots,\delta_4$ are arbitrary constants,
$(\delta_1,\delta_2)\neq(0,0)$ and  $\lambda=1/\alpha_t$.
\end{theorem}
The usual equivalence group~$G^{\sim}_1$  of class~(\ref{vc_mKdV}) is the subgroup of the generalized extended
equivalence group~$\hat G^{\sim}_1$, which is singled out with the condition $\delta_1=\delta_3=0$.

The parameterization of transformations from~$\hat G^{\sim}_1$ by the arbitrary function $\alpha(t)$
allows us to simplify the group classification problem for class~(\ref{vc_mKdV}) via reducing the number of arbitrary elements.
For example, we can gauge arbitrary elements via setting either $h=0$ or $g=1$.
Thus, the gauge $h=0$ can be made by the equivalence transformation
\begin{equation}\label{gauge_h=0}
\hat t=\int e^{-\int h(t)\, dt}dt,\quad \hat x=x, \quad
\hat u=e^{\int h(t)\, dt}u,
\end{equation}
that connects equation~(\ref{vc_mKdV}) with the equation
$\hat u_{\hat t}+\hat u\hat u_{\hat x}+\hat g(\hat t){\hat u}_{\hat x\hat x\hat x}=0.$
The new arbitrary element $\hat g$ is expressed via $g$ and $h$ in the following way:
\[
\hat g(\hat t)=e^{\int h(t)\, dt}g(t).
\]

This is why without loss of generality we can restrict the study to the class
\begin{equation}\label{vc_mKdV_h=0}
u_t+uu_{x}+g(t)u_{xxx}=0,
\end{equation}
since all results on symmetries and exact solutions for this class can be extended to
class~(\ref{vc_mKdV}) with transformations of the form~(\ref{gauge_h=0}).

The equivalence group for class~(\ref{vc_mKdV_h=0}) can be obtained
from Theorem 1 by setting
$\tilde h=h=0$. Note that class~(\ref{vc_mKdV_h=0}) is normalized in the usual sense.

\begin{theorem}[\cite{Popovych&Vaneeva2010}] The equivalence group~$G^{\sim}_0$ of class~(\ref{vc_mKdV_h=0})
is formed by the transformations
\begin{eqnarray*}
&\displaystyle\tilde t=\frac{at+b}{ct+d},\quad
\tilde x=\frac{e_2x+e_1t+e_0}{ct+d},\\
&\displaystyle\tilde u=\frac{e_2(ct+d)u-e_2cx-e_0c+e_1d}\varepsilon,\quad
\tilde g=\frac{e_2{}^3}{ct+d}\frac g\varepsilon,
\end{eqnarray*}
where $a$, $b$, $c$, $d$, $e_0$, $e_1$ and $e_2$ are arbitrary constants with $\varepsilon=ad-bc\ne0$ and $e_2\ne0$,
the tuple  $(a,b,c,d,e_0,e_1,e_2)$ is defined up to nonzero multiplier
and hence without loss of generality we can assume that $\varepsilon=\pm1$.
\end{theorem}


\section{Lie symmetries}
The group classification of class~(\ref{vc_mKdV_h=0}) up to $G_0^\sim$-equivalence is carried out in~\cite{Popovych&Vaneeva2010} in the framework of classical approach~\cite{Olver1986,Ovsiannikov1982}.
The result reads as follows.

The kernel of the maximal Lie invariance algebras of equations from class~(\ref{vc_mKdV_h=0})
coincides with the two-dimensional algebra $\langle\partial_x,\, t\partial_x+\partial_u\rangle$.
All possible $G_0^\sim$-inequiva\-lent cases of extension of the maximal Lie invariance algebras are exhausted
by the cases 1--4 of Table~1.
\begin{table}\centering
\caption{The group classification of the class $u_t+uu_{x}+g\,u_{xxx}=0$, $g\neq0$}
\label{Vaneeva:table1}
\begin{tabular}{@{\,\,}c@{\,\,}@{\,\,}c@{\,\,}@{\,\,}l@{\,\,}}
\hline\noalign{\smallskip}
N&$g(t)$&\hfil Basis of $A^{\max}$ \\
\noalign{\smallskip}\svhline\noalign{\smallskip}
 0&$\forall$&$\partial_x,\quad t\partial_x+\partial_u$\\[0.5ex]
1&
$ t^n$&$\partial_x,\quad t\partial_x+\partial_u,\quad  3t\partial_t+(n+1)x\partial_x+(n-2)u\partial_u$\\[0.5ex]
2&$ e^{t}$&
$\partial_x,\quad t\partial_x+\partial_u,\quad  3\partial_t+x\partial_x+u\partial_u$\\[0.5ex]
3&$ e^{\delta\arctan t}\sqrt{t^2+1}$&
$\partial_x,\quad t\partial_x+\partial_u,\quad  3(t^2+1)\partial_t+(3t+\delta)x\partial_x+((-3t+\delta)u+3x)\partial_u$\\[0.5ex]
4&$1$&
$\partial_x,\quad t\partial_x+\partial_u,\quad 3t\partial_t+x\partial_x-2u\partial_u,\quad \partial_t$\\
\noalign{\smallskip}\hline\noalign{\smallskip}
\end{tabular}

Here $n, \delta$ are arbitrary constants, $n\geq1/2$, $n\neq1$, $\delta\geq 0\ {\rm mod}\  G^\sim_0.$
\end{table}

For any equation from class~(\ref{vc_mKdV}) there exists
an imaged equation in class~(\ref{vc_mKdV_h=0}) with respect to transformation~(\ref{gauge_h=0}).
The equivalence group
$G_0^\sim$ of class~(\ref{vc_mKdV_h=0}) is induced by the equivalence group~$\hat G^\sim_1$ of class~(\ref{vc_mKdV})
which, in turn, is induced by the equivalence group~$G^\sim$ of class~(\ref{EqvcmKdV}).
These guarantee that
Table~\ref{Vaneeva:table1} presents also the group classification list for class~(\ref{vc_mKdV}) up to $\hat G^\sim_1$-equivalence
(resp. for the class~(\ref{EqvcmKdV}) up to $G^\sim$-equivalence).
As all of the above classes are normalized, we can state that we obtain Lie symmetry classifications of these classes
up to general point equivalence.
This leads to the following assertion.
\begin{corollary}
An equation from class~(\ref{vc_mKdV}) (resp.~(\ref{EqvcmKdV})) admits
a four-dimensional Lie invariance algebra if and only if
it is reduced by a point transformation to constant coefficient KdV equation, i.e., if and only if condition~(\ref{EqvcKdVEquivToKdV2}) (resp.~(\ref{EqvcKdVEquivToKdV})) holds.
\end{corollary}

To derive the group classification of class~(\ref{vc_mKdV}) which is not simplified by equivalence transformations,
we first apply transformations from the group $G_0^\sim$ to the classification list presented in Table~\ref{Vaneeva:table1} and obtain the following extended list:

\medskip

0. arbitrary $\hat g\colon$  $\langle\partial_{\hat x},\  \hat t\partial_{\hat x}+\partial_{\hat u}\rangle$;

\smallskip

1. $\displaystyle \hat g=c_0(a\,\hat t+b)^n(c\,\hat t+d)^{1-n}$,
$n\ne0,1$:\quad$\langle \partial_{\hat x},\ \hat t\partial_{\hat x}+
  \partial_{\hat u},\ X_3\rangle$,\quad where
\begin{eqnarray*}&X_3= 3(a\,\hat t+b)(c\,\hat t+d)\partial_{\hat t}+
  \left(3ac\hat t+ad(n+1)+bc(2-n)\right)\hat x\partial_{\hat x}+\\
  &\left[3ac\hat x-(3ac\hat t+ad(2-n)+bc(n+1))\hat u\right]\partial_{\hat u};
\end{eqnarray*}

2. $\displaystyle \hat g=c_0(c\,\hat t+d)\exp\left(\frac{a\,\hat t+b}{c\,\hat t+d}\right)$:\quad
$\langle \partial_{\hat x},\ \hat t\partial_{\hat x}+
  \partial_{\hat u},\ X_3\rangle$,\quad where
\[X_3=3(c\,\hat t+d)^2\partial_{\hat t}+\left(3c(c\hat t+d)+\varepsilon\right)\hat x\partial_{\hat x}+\\
 \left[3c^2\hat x+ (\varepsilon-3c(c\hat t+d))\hat u\right]\partial_{\hat u};
\]

3. $\displaystyle
\hat g=c_0e^{\delta\arctan\left(\frac{a\,\hat t+b}{c\,\hat t+d}\right)}
\sqrt{(a\,\hat t+b)^2+(c\,\hat t+d)^2}$:\quad $\langle \partial_{\hat x},\ \hat t\partial_{\hat x}+  \partial_{\hat u},\ X_3\rangle$,\quad where
\begin{eqnarray*}&X_3= 3\left((a\,\hat t+b)^2+(c\,\hat t+d)^2\right)\partial_{\hat t}+
  \left(3a(a\hat t+b)+3c(c\hat t+d)+\varepsilon\delta\right)\hat x\partial_{\hat x}+\\
  &\left(3(a^2+c^2)\hat x-(3a(a\hat t+b)+3c(c\hat t+d)-\varepsilon\delta)\hat u\right)\partial_{\hat u};
\end{eqnarray*}

4a. $\hat g=c_0$: \quad
$\langle\partial_{\hat x},\ \hat t\partial_{\hat x}+
  \partial_{\hat u},\ \partial_{\hat t},\,3\hat t\partial_{\hat t}+\hat x\partial_{\hat x}-2\hat u\partial_{\hat u}\rangle;$

\smallskip

4b. $\hat g=c\hat t+d$, $c\ne0$: \quad
$\langle\partial_{\hat x},\ \hat t\partial_{\hat x}+
  \partial_{\hat u},\ 3(c\hat t+d)\partial_{\hat t}+2c\hat x\partial_{\hat x}-c\hat u\partial_{\hat u},\ X_4\rangle$,\quad  where
\[
X_4= (c\hat t+d)^2\partial_{\hat t}+c(c\hat t+d)\hat x\partial_{\hat x}+
  c(c\hat x-(c\hat t+d)\hat u)\partial_{\hat u}.
\]
Here $c_0$, $a$, $b$, $c$, $d$ and $\delta$ are arbitrary constants, $(a^2+b^2)(c^2+d^2)\ne0$, $\varepsilon=ad-bc,$ $c_0\neq0$.

Then we find preimages of equations from the class $\hat  u_{\hat  t}+\hat  u\hat  u_{\hat  x}+\hat  g(\hat  t){\hat  u}_{\hat  x\hat  x\hat  x}=0$
with arbitrary elements collected in the above list with respect to transformation~(\ref{gauge_h=0}).
The last step is to transform basis operators of the corresponding Lie symmetry algebras. The results are presented in Table~2.
\begin{table}
\caption{The group classification of the class $u_t+uu_{x}+gu_{xxx}+hu=0$, $g\neq0$}
\label{Vaneeva:table2}
\begin{tabular}{@{\,\,}c@{\,\,}@{\,\,}c@{\,\,}@{\,\,}c@{\,\,}@{\,\,}l@{\,\,}}
\hline\noalign{\medskip}
N&$h(t)$&$g(t)$&\hfil Basis of $A^{\max}$ \\
\noalign{\smallskip}\svhline\noalign{\medskip}
 0&$\forall$&$\forall$&$\partial_x,\ T\partial_x+T_t\partial_u$\\[1ex]
1&$\forall$&$c_0T_t(aT+b)^n(cT+d)^{1-n}$
&$ \partial_x,\ T\partial_x+T_t\partial_u,\  3T_t^{-1}(aT+b)(cT+d)\partial_t+\bigl[3acT+$\\[0.5ex]
&&&$ad(n+1)+bc(2-n)\bigr]x\partial_x+\Bigl(3acxT_t-\bigl[3acT+$\\[0.5ex]
&&&$3hT_t^{-1}(aT+b)(cT+d)+bc(n+1)+ad(2-n)\bigr]u\Bigr)\partial_u$\\[1ex]
2&$\forall$&$c_0T_t(cT+d)\exp\left(\frac{aT+b}{cT+d}\right)$&
$ \partial_x,\ T\partial_x+T_t\partial_u,\  3T_t^{-1}(cT+d)^2\partial_t+(3c(cT+d)+\varepsilon)x\partial_x+$\\[0.5ex]
&&&$\left[3c^2xT_t+\left(\varepsilon-3(cT+d)(c+h(cT+d)T_t^{-1})\right)u\right]\partial_u$\\[1ex]
3&$\forall$&$c_0T_te^{\delta\arctan\left(\frac{aT+b}{cT+d}\right)}G(t)$&
$ \partial_x,\ T\partial_x+T_t\partial_u,\ 3T_t^{-1}G^2\partial_t+$\\[0.5ex]
&&&$\bigl[3a(aT+b)+3c(cT+d)+\varepsilon\delta\bigr] x\partial_{ x}+\bigl[3(a^2+c^2)xT_t-$\\[0.5ex]
&&&$\bigl(3a(aT+b)+3c(cT+d)-\varepsilon\delta+3hT_t^{-1}G^2\bigr)u\bigr]\partial_u$\\[1ex]
4a&$\forall$&$c_0T_t$&
$ \partial_x,\ T\partial_x+T_t\partial_u,\   T_t^{-1}(\partial_t-hu\partial_u),$\\[0.5ex]
&&&$ 3TT_t^{-1}\partial_t+x\partial_x-(2+3TT_t^{-1}h)u\partial_u$\\[1ex]
4b&$\forall$&$(cT+d)T_t$&
$ \partial_x,\ T\partial_x+T_t\partial_u,\  T_t^{-1}(cT+d)^2\partial_t+c(cT+d)x\partial_x+$\\[0.5ex]
&&&$[c^2xT_t-(cT+d)(c+T_t^{-1}(cT+d)h)u]\partial_u,$\\[0.5ex]
&&&$  3T_t^{-1}(cT+d)\partial_t+2cx\partial_x-(c+3T_t^{-1}(cT+d)h)u\partial_u$\\
\noalign{\smallskip}\hline\noalign{\smallskip}
\end{tabular}

Here $T=\int e^{-\int h(t)\, dt}dt$, $T_t=e^{-\int h(t)\, dt}$, $G=\sqrt{(aT+b)^2+(cT+d)^2}$; $n $ $c_0$, $a$, $b$, $c$, $d$ and $\delta$ are arbitrary constants, $(a^2+b^2)(c^2+d^2)\ne0$, $\varepsilon=ad-bc,$ $c_0\neq0$, $n\neq0,1$. In the case (4b) $c\neq0$.
\end{table}

It is easy to see that Table 2 includes all cases presented in~\cite{john10b} as particular cases.

\section{Generation of exact solutions}
A number of recent papers concern the construction of exact solutions to different classes of KdV- or mKdV-like equations
using, e.g., such methods as ``generalized $(G'/G)$-expansion method'', ``Exp-function method'',
``Jacobi elliptic function expansion method'', etc. A number of references are presented in~\cite{Popovych&Vaneeva2010}.
Almost in all cases exact solutions were constructed only for equations which are reducible to the standard
KdV or mKdV equations by point transformations and usually these were only solutions similar to the well-known one-soliton solution.
In this section we show that the usage of equivalence transformations allows one to obtain more results in a simpler way.
This approach is used also in~\cite{Tang&Zhao&Huang&Lou2009}.

The $N$-soliton solution of the KdV equation in the canonical form
\begin{equation}\label{canonical_KdV}
U_t-6UU_{x}+U_{xxx}=0
\end{equation}
was constructed as early as in the seventies by Hirota~\cite{Polyanin&Zaitsev}.
The  two-soliton solution of equation~(\ref{canonical_KdV}) has the form
\begin{equation}\label{sol2soliton}
U=-2\frac{\partial^2}{\partial x^2}\ln\left(1+b_1e^{\theta_1}+b_2e^{\theta_2}+Ab_1b_2e^{\theta_1+\theta_2}\right),
\end{equation}
where $a_i, b_i$ are arbitrary constants, $\theta_i=a_ix-a_i^3t,$ $i=1,2;$ $ A=\left(\frac{a_1-a_2}{a_1+a_2}\right)^2$.

Combining the simple transformation $\hat u=-6 U$ that connects~(\ref{canonical_KdV})
with the KdV equation of the form
\begin{equation}\label{KdV_canonical}
\hat u_{\hat t}+{\hat u}\hat u_{\hat x}+\hat u_{\hat x\hat x\hat x}=0
\end{equation}
and transformation~(\ref{gauge_h=0}), we obtain the formula
\[\textstyle  u=-6e^{-\int h(t)dt}\,U\left(\int e^{-\int h(t)\, dt}dt,\,x\right)\]
for generation of exact solutions for the equations of the general form
\begin{equation}\label{KdV_canonical_preimage}
u_t+uu_{x}+e^{-\int h(t)\, dt}u_{xxx}+h(t)u=0.
\end{equation}
These equations are preimages of~(\ref{KdV_canonical}) with respect to transformation~(\ref{gauge_h=0}).
Here $h$ is an arbitrary nonvanishing smooth function of the variable~$t$.

The two-soliton solution~(\ref{sol2soliton}) leads to the following solution of~(\ref{KdV_canonical_preimage})
\begin{equation}
u=12e^{-\int h(t)dt}\frac{\partial^2}{\partial x^2}\ln\left(1+b_1e^{\theta_1}+b_2e^{\theta_2}+Ab_1b_2e^{\theta_1+\theta_2}\right),
\end{equation}
where $a_i, b_i$ are arbitrary constants, $\theta_i=a_ix-a_i^3\int e^{-\int h(t)\, dt}dt,$ $i=1,2;$ $ A=\left(\frac{a_1-a_2}{a_1+a_2}\right)^2$.
In a similar way one can construct $N$-soliton, rational and other types of solutions for equations from class~(\ref{KdV_canonical_preimage}) using known solutions of classical KdV equation.

\section{Conclusion}
In this paper group classification problem for class~(\ref{vc_mKdV}) is carried out with respect
to the corresponding equivalence group using equivalence based approach. Using the normalization property it is proved that this classification coincides
with the one carried out up to general point equivalence.
The classification list  extended by equivalence transformations is also presented. Such list is convenient for further applications.

It is shown that the usage of equivalence groups is a crucial point for exhaustive solution of the problem.
Moreover, equivalence transformations allow one to construct exact solutions of different types in a much easier way than by direct solving.
These transformations can also be utilized to obtain conservation laws, Lax pairs and other related objects
for equations reducible to well-known equations of mathematical physics by point transformations without direct calculations.

\begin{acknowledgement}
The author thanks the Organizing Committee  and especially Prof. Vladimir Dobrev for hospitality and giving an opportunity to give a talk.
Her participation in the Workshop was supported by the Abdus Salam International Centre for Theoretical Physics.
The author is also grateful to Prof. Roman Popovych for useful discussions and valuable comments.
\end{acknowledgement}


\begin{thebibliography}{99}
\footnotesize\itemsep=-.3ex


\bibitem{Gungor&Lahno&Zhdanov2004}
 F. G\"ung\"or, V.I. Lahno, R.Z. Zhdanov,
 J. Math. Phys.    {\bf 45} (2004) 2280--2313.


\bibitem{IPS2007a}
N.M. Ivanova, R.O. Popovych, C. Sophocleous,
Lobachevskii J. Math. {\bf 31} (2010) 100--122.



\bibitem{john10b}A.G. Johnpillai, C.M. Khalique, Appl. Math. Comput. {\bf 216} (2010) 3761--3771.




\bibitem{Magadeev1993}
 B.A. Magadeev,
 Algebra i Analiz {\bf 5} (1993) 141--156 (in Russian);
translation in  St. Petersburg Math. J. {\bf 5} (1994) 345--359.

\bibitem{Olver1986}P. Olver, {\it Applications of Lie groups to differential equations},
(Springer-Verlag, New York, 1986).


\bibitem{Ovsiannikov1982}
L.V. Ovsiannikov, {\it Group analysis of differential equations},
(Academic Press, New York, 1982).

\bibitem{Polyanin&Zaitsev}
A.D. Polyanin, V.F. Zaitsev, {\it Handbook of Nonlinear Partial Differential Equations}, (Chapman \&
Hall/CRC Press, Boca Raton, 2004).

\bibitem{Popovych2006c}
 R.O. Popovych,
in {\it Collection of Works of Institute of Mathematics} (Institute of Mathematics, Kyiv, Ukraine) {\bf 3}, no.~2 (2006) 239--254.

\bibitem{Popovych&Kunzinger&Eshraghi2010}
R.O. Popovych, M. Kunzinger, H. Eshraghi,
{\it Acta Appl. Math.} {\bf 109} (2010) 315--359.

\bibitem{Popovych&Vaneeva2010}
R.O. Popovych, O.O. Vaneeva,  Commun. Nonlinear Sci. Numer. Simulat. {\bf 15} (2010) 3887--3899.



\bibitem{Tang&Zhao&Huang&Lou2009}
X.Y. Tang, J. Zhao, F. Huang, S.Y. Lou,
 Studies in Appl. Math. {\bf 122} (2009) 295--304.

\bibitem{Vaneeva2012}
O.O. Vaneeva, Commun. Nonlinear Sci. Numer. Simulat. {\bf 17} (2012) 611--618.

\bibitem{VJPS2007}
O.O. Vaneeva, A.G. Johnpillai, R.O. Popovych, C. Sophocleous,
J. Math. Anal. Appl. {\bf 330} (2007) 1363--1386.

\bibitem{VPS_2009}
O.O. Vaneeva, R.O. Popovych, C. Sophocleous,  Acta Appl. Math. {\bf 106} (2009) 1--46.
\end{thebibliography}
\end{document}